\documentclass[]{aa} 

\usepackage{graphicx}
\usepackage{txfonts}
\usepackage{natbib}
\usepackage{longtable}

\begin{document}

\title{Reliability of NH$_3$ as the temperature probe of cold cloud cores}

\author{M.     Juvela\inst{1},
        J.     Harju\inst{2,1},
        N.     Ysard\inst{1},
        T.     Lunttila\inst{1},
        }

\institute{
Department of Physics, P.O.Box 64, FI-00014, University of Helsinki,
Finland, {\em mika.juvela@helsinki.fi}
\and
Finnish Centre for Astronomy with ESO (FINCA)
}

\authorrunning{M. Juvela et al.}

\date{Received September 15, 1996; accepted March 16, 1997}

\abstract
{
The temperature is a central parameter affecting the chemical and
physical properties of dense cores of interstellar clouds and their
potential evolution towards star formation. The chemistry and the dust
properties are temperature dependent and, therefore, interpretation of
any observation requires the knowledge of the temperature and its
variations. Direct measurement of the gas kinetic temperature is
possible with molecular line spectroscopy, the ammonia molecule,
NH$_3$, being the most commonly used tracer.
}
{
We want to determine the accuracy of the temperature estimates derived
from ammonia spectra. The normal interpretation of NH$_3$ observations
assumes that all the hyperfine line components are tracing the same
volume of gas. However, in the case of strong temperature gradients
they may be sensitive to different layers and this could cause errors in
the optical depth and gas temperature estimates.
}
{
We examine a series of spherically symmetric cloud models, 1.0 and
0.5\,$M_{\sun}$ Bonnor-Ebert spheres, with different radial
temperature profiles. We calculate synthetic NH$_3$ spectra and
compare the derived column densities and temperatures to the actual
values in the models.
}
{
For high signal-to-noise observations, the estimated gas kinetic
temperatures are within $\sim$0.3\,K of the real mass averaged
temperature and the column densities are correct to within $\sim$10\%.
When the S/N ratio of the (2,2) spectrum decreases below 10, the
temperature errors are of the order of 1\,K but without a significant
bias. Only when the density of the models is increased by a factor of
a few, the results begin to show significant bias because of the
saturation of the (1,1) main group.
}
{
The ammonia spectra are found to be a reliable tracer of the real mass
averaged gas temperature. Because the radial temperature profiles of
the cores are not well constrained, the central temperature could
still be different from this value. If the cores are optically very
thick, there are no longer guarantees of the accuracy of the
estimates.
}
\keywords{
ISM: clouds -- ISM: molecules -- Radio lines: ISM -- Stars: formation
-- radiative transfer
}

\maketitle
%

\section{Introduction}

The dense molecular cloud cores are of particular interest because
their collapse can lead to the formation of new stars. To understand
the star formation process, one needs to understand the physical and
chemical properties of the cores. The temperature of the cores,
especially as a function of position, is difficult to measure and yet
it affects everything from core stability to chemistry and dust
evolution.

Dust and gas become thermally coupled only at high densities,
$n$(H$_2$)$\sim 10^5$\,cm$^{-3}$ and above \citep{Goldsmith2001}.
Because the continuum surface brightness is strongly affected
by the warm outer cloud layers, observations of the dust emission
are not ideal to determine the gas temperature at the centre of the
dense cores. The widths of molecular lines can sometimes be used
to derive stringent limits on the gas temperature
\citep[e.g.,][]{Harju2008} but the value of this method is limited
by the presence of turbulent velocity component. 

In observations of most molecules, it is difficult to separate the
effects of density, optical depth, and temperature. The situation
is different for ammonia, NH$_3$, whose rotational energy is
characterised by the quantum numbers ($J$,$K$). Because the dipole
transitions between the different $K$ ladders are forbidden, their
relative populations depend only on collisions and thus measure the
gas kinetic temperature. Each rotational state ($J$,$K$) is further
divided to inversion doublets. The inversion transitions (1,1) and
(2,2) are at 23.7\,GHz and 23.72\,GHz respectively, and are easily
observable from the ground. The inversion transitions also have a
hyperfine structure. For example the (1,1) spectrum contains 18
components from which at least five separate groups of lines are
usually discernible. The intensity ratio of the hyperfine
components provides a convenient tool to estimate the line optical
depths. Because of these properties, ammonia is a common tool for
gas temperature measurements, especially in the case of dense cores
\citep{Ho1983}. The high critical density ($\sim2\cdot
10^3$\,cm$^{-3}$ for the (1,1) transition as calculated from
the ratio of the spontaneous and collisional de-excitation rates at
the temperature of 15\,K) guarantees that one is measuring
temperature of dense gas. On the other hand, NH$_3$ is resilient
against depletion and the abundance is observed to drop only in the
very densest and coldest cores \citep{BerginTafalla2007}.

The analysis of NH$_3$ is based on the assumption that the ($J$,$K$)
transitions and their different hyperfine components are all tracing
the same physical region. Because of the different optical depths, the
weaker hyperfine components could be more sensitive to deeper layers
in the cloud. The effect should become noticeable if the main lines
are moderately optically thick and the source has strong temperature
gradients.  

The temperature structure of quiescent cores can, in principle, be
calculated from the balance of the various heating and cooling
mechanisms \cite[e.g.][]{Goldsmith2001, Galli2002}. The gas within the
cores is heated by the cosmic rays and the external stellar radiation
field and it is cooled down by the line radiation from a number of
atomic and molecular species. Depending on the temperature difference
between the gas and the dust, the gas-dust collisions can result in
either heating or cooling of the gas.
\cite{JuvelaYsard2011} discussed the range of possible temperature
profiles in dense cores, emphasising the uncertainties associated with
many of the contributing factors (e.g., the external radiation field,
the gas-dust coupling, and the abundances of the species responsible
for the main cooling lines).
In this paper we use the models of \cite{JuvelaYsard2011} to examine
the reliability of the ammonia temperature estimates in the presence
of radial temperature variations.

The content of the paper is the following. In
Sect.~\ref{sect:methods} we present the methods concerning the core
models, the calculation of synthetic ammonia spectra, and the
derivation of temperature and column density estimates. The results
are presented in Sect.~\ref{sect:results}, where we compare the
derived estimates to the corresponding actual values in the models, in
particular studying the effects of observational noise and increased
column densities. The
final conclusions are presented in Sect.~\ref{sect:conclusions}.

\section{Methods}  \label{sect:methods}

\subsection{Core models} \label{sect:models}

We use the cloud models discussed by \cite{JuvelaYsard2011}. These are
0.5\,$M_{\sun}$ and 1.0\,$M_{\sun}$ Bonnor-Ebert spheres ($\xi$=6.5,
$T_{\rm kin}$=10\,K) for which the gas temperatures were calculated
under different hypotheses. This resulted in a set of radial
temperature profiles with varying shapes, the temperature being of the
order of 10\,K. 
While the dust temperature decreases inwards in starless cores, the
situation can be different for the gas temperature because of the
efficient molecular line cooling in the outer part.
In the models, the gas temperature profiles were mostly increasing
towards the centre. When the photoelectric heating was included, the
situation is reversed and the surface temperature is raised close to
20\,K.

The calculations follow \cite{Goldsmith2001} apart from the
fact that the cooling rates due to line emission are obtained with a
Monte Carlo radiative transfer program. The cooling rates are
calculated directly for $^{12}$CO, $^{13}$CO, C$^{18}$O, C, CS, and
o-H$_2$O. The rates of the last two species are multiplied by 10.0 and
2.0, respectively, to take into account the contribution of other
species. We examine the following cases without the photoelectric
heating: the default model (`{\em def}'), the low abundance model
($\chi$), the weak gas-dust coupling model ($\Lambda_{\rm gd}$), the
small velocity dispersion model ($\sigma_{\rm v}$), and the large
velocity gradient model ($v$). The names refer to the temperature
calculations in \cite{JuvelaYsard2011}. 

The $def$-model was obtained assuming $\sigma_{\rm
v}$=1.0\,km\,s$^{-1}$ line widths (Doppler width) and constant
abundances in the calculation of the cooling lines, no large scale
velocity field, and a coupling between the gas and dust temperatures
that followed that presented in \cite{Goldsmith2001}. In the
$\sigma_{\rm v}$-model, the turbulent linewidth changes as a step
function from 1.0\,km\,s$^{-1}$ in the outer part, outside the half
radius, to 0.1\,km\,s$^{-1}$ within the inner core. The $v$-model is
the only one with a large scale velocity field. It has radial infall
velocities that increase linearly from 0\,km\,s$^{-1}$ at the surface
to 1.0\,km\,s$^{-1}$ in the centre. The name of the low abundance
model ($\chi$) refers to the calculations of the gas kinetic
temperature where the abundance of the cooling species was assumed to
be lower at the centre of the core. This was implemented as a
step function with a factor of ten difference between the inner and
outer parts. However, in the modelling of the NH$_3$ spectra, the
fractional abundance of para-NH$_{3}$ is always kept at a constant
value of 10$^{-8}$ \citep[e.g.][]{Hotzel2001, Tafalla2004}. This
applies also to the $\chi$-model. The photoelectric heating is
included only in one model (model $PEH$) where the model cloud is
assumed to be illuminated by an unattenuated interstellar radiation
field \citep{Mathis1983}. For the details of the calculations of the
$T_{\rm kin}$ profiles, see \cite{JuvelaYsard2011}.

In addition to the above described models, we consider cases where the
kinetic temperatures are the same but the turbulent line width is
decreased by a factor of ten in the NH$_3$ calculations. In
particular, in the central part of the $\sigma_{\rm V}$-model the
linewidth then becomes practically thermal. 
We use the \cite{JuvelaYsard2011} results only as examples of the
possible shapes of the temperature profiles and the purpose of the
present paper is to probe their effect on the analysis of ammonia
spectra at a general level.

%
                                                 
\begin{table}
\caption{The main core models.}
\centering
\begin{tabular}{lll}
\hline\hline
Model                &  Modification                  &  $T_{\rm kin}$ ranges
of the 0.5 \\
                     &                                       &  and
                     1.0\,M$_{\sun}$ models (K) \\
\hline
Default              &  --                                   &  5.3-9.3, 4.6-8.4    \\
$\chi$               &  $\chi= (0.1-1.0)$\,\,$\chi^{\rm default}$  &  5.3-11.0, 4.8-11.0  \\                        
$\Lambda_{\rm g, d}$ &  $\Lambda_{\rm g, d}$=(0.3-1.0)\,\,$\Lambda_{\rm g, d}^{\rm default}$
                     &  6.4-10.0, 5.3-8.5 \\
$\sigma_{\rm v}$     &  $\sigma_{\rm v}$=(0.1-1.0)\,\,$\sigma_{\rm v}^{\rm default}$ &
                        5.3-10.9, 4.8-11.4 \\
$v_{\rm radial}$     &  0-1.0\,km\,s$^{-1}$   &   5.3-9.1, 4.8-8.3 \\
PEH                  &  include $\Gamma(PEH)$ for    &
                        9.2-14.7, 9.2-19.1 \\
                     &  $\zeta$=3$\cdot 10^{17}$\,s$^{-1}$ & \\
\hline
\end{tabular}
\end{table}

\subsection{Synthetic spectra} \label{sect:rt}

For each model, synthetic para-NH$_3$ (1,1) and (2,2) spectra were
obtained with the help of radiative transfer calculations.
The non-LTE radiative transfer problem is solved with a Monte Carlo
program \citep{Juvela1997}. The the model clouds are divided into 100
shells. To get better resolution in the outer regions where the
excitation conditions can change more rapidly, the thickness of the
shells was decreased towards the surface. The radius of the innermost
sphere and the thickness of the outermost shell are, respectively,
$\sim$6\% and 0.7\% of the radius of the model cloud.

The ammonia spectra are calculated following the prescription
presented in \cite{Keto2004}. The non-LTE populations of the ($J$,$K$)
states of para-NH$_3$ are obtained from the radiative transfer
calculations where, within each state, an LTE distribution is assumed
between the hyperfine levels (i.e., distribution according to the
statistical weights of the transitions). In the radiative transfer
calculations, we take into account the hyperfine components of the
(1,1) and (2,2) lines. These consist of 18 and 21 components,
respectively. In the final synthetic (1,1) and (2,2) spectra, many of
these components are blended together. The molecular parameters for
the ($J$,$K$) lines are from the Lamda database \citep{Schroier2005}
and the frequencies and relative intensities of the hyperfine lines
were taken from \cite{Kukolich1967}. 

\begin{figure}
\centering
\includegraphics[width=8cm]{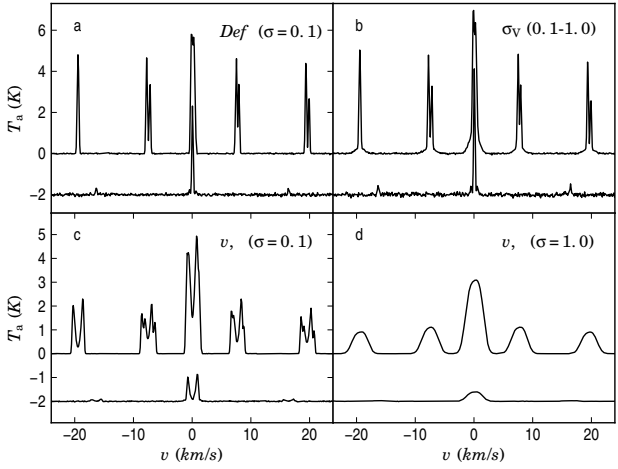}
\caption{
Examples of simulated ammonia spectra. Each frame shows the (1,1) (the
upper line) and the (2,2) spectral profiles. For the plotting, the
latter have been scaled by a factor of 3 and shifted by -2\,K. The
text in the upper right hand corner indicates the model in question
and, in parentheses, the turbulent line width used in the calculation
of the NH$_3$ spectra.
}
\label{fig:spectra}%
\end{figure}

The synthetic spectra are calculated for a pencil beam (i.e.,
for a single line-of-sight through the centre of the core) and
averaged over a Gaussian beam with the FWHM equal to the core
radius. Because the excitation of the molecules varies along the
line of sight and the hyperfine components have different optical
depths, the observed intensity ratios between hyperfine components
can vary depending on the optical depth and temperature structure
of the model. Figure~\ref{fig:spectra} shows examples of the
simulated spectra.

\subsection{Analysis of the spectra} \label{sect:analysis}


To derive estimates of the column densities and gas kinetic
temperatures, we subject the synthetic NH$_3(1,1)$ and NH$_3(2,2)$
spectra to standard analysis as described in \cite{Ho1983},
\cite{WalmsleyUngerechts1983}, and \cite{Ungerechts1986}. The
hyperfine structure of the NH$_3(1,1)$ line is fitted using a $\chi^2$
routine assuming a gaussian velocity distribution and that the
different hyperfine states are populated according to LTE. The fit
results are used to calculate the optical thickness profile $\tau(v)$.
The peak value $\tau_{\rm peak}$, and the brightness temperature in the
corresponding channel (usually coincident with the intensity maximum
$T_{\rm B,peak}$) are substituted in the antenna equation to derive
the excitation temperature, $T_{\rm ex}$, of the transition.  The
integrated optical thickness, $\int \tau(v) dv$, where $v$ is the
radial velocity, and $T_{\rm ex}$ give the column density in the upper
transition level, $N_{\rm u}(1,1)$, from the formula
$$
N_{\rm u}(1,1) \;=\; \frac{8\pi}{\lambda_{11}^3} \; \frac{1}{A_{\rm
ul}} \; 
F(T_{11},T_{\rm ex}) \; \int \, \tau(v) \,dv \; ,
$$
where $A_{\rm ul}$ is the Einstein coefficient for spontaneous
emission, 
$\lambda_{11} = c/\nu_{11}$ is the wavelength of the transition,
$T_{11}=h\nu_{11}/k_{\rm B}$, and the function $F(T_{0},T)$ is defined
by 
$$
F(T_{0},T) \equiv \frac{1}{e^{T_{0}/T}-1} \; .
$$
Adding together the upper and lower levels of the inversion
transition, the total column density of ammonia in the (1,1) level 
is
$$
N(1,1) \; = \; N_{\rm u} \; (1 + e^{T_{11}/T_{\rm ex}}) \, .
$$
  
The column density of the (2,2) level, $N(2,2)$, can be derived using
two slightly different methods: a) by estimating $\tau$ and $T_{\rm
ex}$ from the hyperfine fit as in the case of (1,1); or b) by assuming
that $T_{\rm ex}$ is the same as for the $(1,1)$ transition and that
the (2,2) line is optically thin. In the case b the integral $\int
\tau(v) dv$ is estimated from the integrated intensity of the (2,2)
main group using the antenna equation. Usually only the method b is
feasible for observational data of cold clouds because the (2,2)
satellites are very weak and buried in the noise. In the present paper
we use the method b to do the analysis.
  
The rotational temperature, $T_{12}$, is defined with the Boltzmann
equation
$$
\frac{N(2,2)}{N(1,1)} = \frac{5}{3} e^{-41.5/T_{12}} \; .
$$
The kinetic temperature, $T_{\rm kin}$, is estimated using the three
level approximation (including the levels $(J,K)$=(1,1), (2,1) and
(2,2)) as outlined by \cite{WalmsleyUngerechts1983}. 
The temperature dependence of the ratio of the collisional
coefficients $C_{22\rightarrow21}/C_{22\rightarrow11}$ was estimated
adopting the coefficients of \cite{Danby1988}.  The method is the same
as used in \cite{Harju1993}.  The resulting relationship between
$T_{12}$ and $T_{\rm kin}$ is very close to that given in the Appendix
B of \cite{Tafalla2004}.
%
In cold gas, the sum of the column densities of the (1,1) and (2,2)
levels is almost equal to the total column density of {\sl
para}-NH$_3$ as the populations of the higher rotational levels are
negligible. We calculate these column density estimates from the
synthetic observations and compare with the corresponding true values
in the model clouds.

\section{Results}  \label{sect:results}

\subsection{The basic models}

The derived {\sl para}-NH$_3$ column densities and the gas kinetic
temperatures were compared to the corresponding true values that were
calculated as the mass averaged quantities in the model clouds,
averaged over the same beam as what was used in the calculation of the
synthetic ammonia spectra. 

\begin{figure}
\centering
\includegraphics[width=8cm]{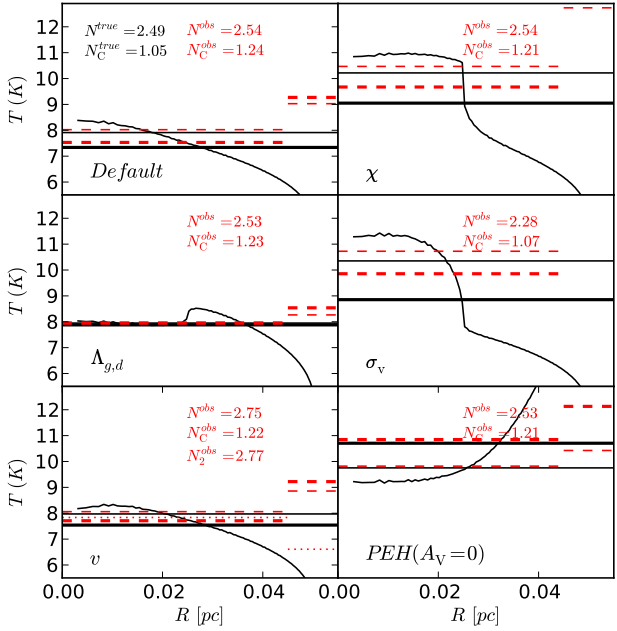}
\caption{
The radial temperature profiles (solid black curves), the estimated
temperatures (dashed red lines), and the mass averaged true
temperatures (black horizontal lines) for the one solar mass
models. The thin lines correspond to the data on a single
line-of-sight through the cloud centre and the thick lines to data
averaged with a gaussian beam.
In the right hand part of each frame, the difference between the
estimated and the real mass averaged $T_{\rm kin}$ has been scaled
by a factor of ten for better visibility.
The column densities derived from the observations are indicated in
each frame (units of $10^{14}$\,cm$^{-2}$), the subscript $C$
denoting the case with beam convolution. The true column densities
are given in the first frame.
For the $v$-model, the dotted line and the lower text entry
indicate the result from a fit with two velocity components (pencil
beam only).
}
\label{fig:M1.0_sigma1.0}%
\end{figure}

\begin{figure}
\centering
\includegraphics[width=8cm]{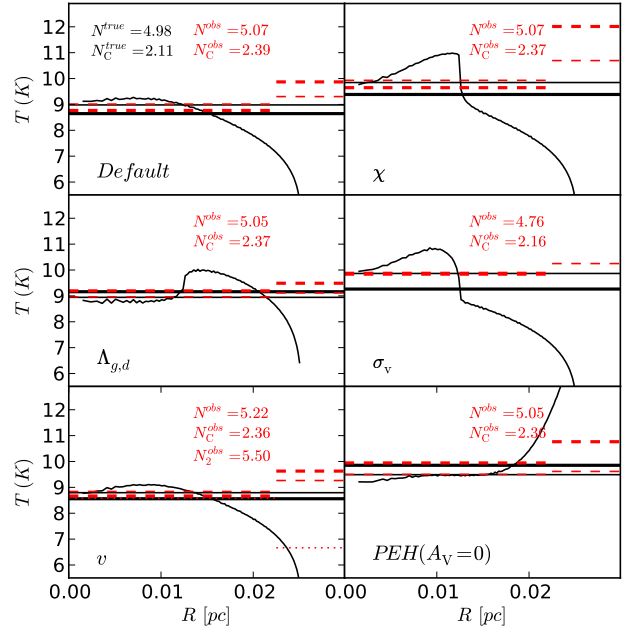}
\caption{
As Fig.~\ref{fig:M1.0_sigma1.0} but for the half solar mass core.
}
\label{fig:M0.5_sigma1.0}%
\end{figure}

\begin{figure}
\centering
\includegraphics[width=8cm]{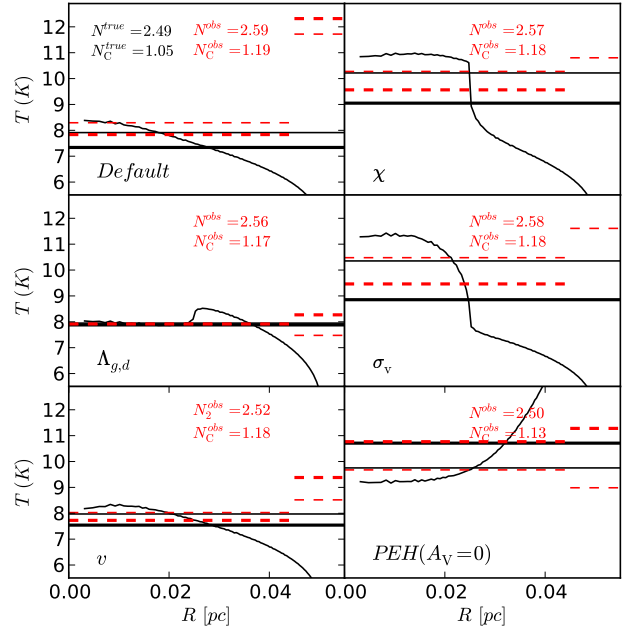}
\caption{
As Fig.~\ref{fig:M1.0_sigma1.0} but with the synthetic ammonia
spectra modelled with a smaller, 0.1\,km\,s$^{-1}$ turbulent
linewidth.
For the model $v$, the pencil beam observations are fitted with
two velocity components and the convolved spectra with a single
velocity component.
}
\label{fig:M1.0_sigma0.1}%
\end{figure}

The main results are presented in
Figs.~\ref{fig:M1.0_sigma1.0}--~\ref{fig:M1.0_sigma0.1}. 
The profiles of the kinetic temperature $T_{\rm kin}$ are shown as
solid black curves. These are a critical input parameter for the
simulation of the ammonia spectra. Without the photoelectric
heating, $T_{\rm kin}$ increases towards the cloud centre. The
central temperature is further increased when, within one half of
the cloud radius, the abundance of the cooling species is decreased
(model $\chi$) or the turbulent linewidth is decreased. In model
$\Lambda_{\rm g, d}$, the gas-dust coupling\footnote{In
\cite{JuvelaYsard2011} the gas-dust coupling $\Lambda_{\rm g, d}$
was calculated using the formulas given in \cite{Goldsmith2001}.
The true heat exchange rate may be higher \cite[see][]{Young2004}.
This would not have a large effect on the temperature in the core
centres where the gas and the dust are well coupled. However, it
would modify somewhat the shape of the $T_{\rm kin}$ profiles.} is
weakened in the inner core, leading to a small decrease in the
temperature because $T_{\rm dust}>T_{\rm gas}$.
We do not claim that all these temperature profiles are likely to
be found in starless cores. However, for the present case the most
important fact is that they probe a wide range of $T_{\rm kin}$
profiles that potentially could result in a bias in the $T_{\rm
kin}$ estimates.
Figure~\ref{fig:M1.0_sigma1.0} shows the $T_{\rm kin}$ estimates for
the 1\,$M_{\sun}$ cloud. The horizontal thin and thick lines indicate
the true mass averaged temperatures for a single line-of-sight through
the centre of the model cloud (thin line) or weighted with a gaussian
beam with a FWHM equal to cloud radius. The dashed lines are
correspondingly the $T_{\rm kin}$ estimates derived from the NH$_3$
spectra observed with a pencil beam or the larger gaussian beam. In
the frames also are given the true and estimated column densities. For
comparison, Figs.~\ref{fig:M0.5_sigma1.0} and ~\ref{fig:M1.0_sigma0.1}
show the results for 0.5\,$M_{\sun}$ models and for 1\,$M_{\sun}$
models with a smaller turbulent linewidth ($\sigma
\le$0.1\,km\,s$^{-1}$).
The temperature estimates derived from the simulated NH$_{3}$ spectra
are correct to within 0.5\,K and the errors of the column density
values are $\sim$10\% or less. This applies to the observations
conducted with a pencil beam and most observations with a gaussian
beam. When one includes the convolution with the large beam, the
temperature values reflect more the values found in the outer part of
the cloud. However, compared to the beam averaged real kinetic
temperature, the values appear to be biased towards the $T_{\rm kin}$
in the cloud centre. The error is largest for the $\sigma_{\rm v}$
model where it can approach 1\,K (Fig.~\ref{fig:M1.0_sigma1.0}).

The spectra of the $v$-model are double peaked because of the
infall motion that at the centre leads to maximum line-of-sight
velocities of $\pm$1\,km\,s$^{-1}$ (see Fig.~\ref{fig:spectra}).  The
redshifted peak is stronger than the blueshifted peak. This is
opposite to the expected behaviour in contracting clouds and is caused
by the fact that the temperature is decreasing towards the core centre
within the innermost $\sim$20\% of the cloud radius. When the spectra
are modelled as the sum of two components, the individual features
described with gaussians, the analysis recovers the correct solution
with good accuracy. In the case of narrow lines ($\sigma
\le$0.1\,km\,s$^{-1}$) a single component fit to the two velocity
components would result in an error of $\sim$1\,K. On the other hand,
when that model is observed with a larger beam, only a single velocity
component remains visible. In Fig.~\ref{fig:M1.0_sigma0.1} the beam
convolved data were fitted with a single gaussian and the pencil beam
observations with two components. In both cases the temperature
estimates remain within $\sim$0.2\,K of the correct value.

\subsection{The influence of noise}

The influence of observational noise was examined by analysing a
series of spectra with progressively higher signal-to-noise ratios.
The S/N ratio is defined as the ratio between the peak intensity of
the (2,2) spectrum and the rms noise per channel. The rms noise added
to the (1,1) spectra was the same in brightness temperature units. In
addition to the default model (marked `$def$'), the models with the
temperature distributions corresponding to the `$\sigma_{\rm v}$' and
`$PEH$' cases were examined, i.e., models with strong inward and
outward temperature gradients. Both the $M=0.5\,M_{\sun}$ and
$M=1.0\,M_{\sun}$ cloud models were used, calculating the ammonia
spectra with either 1.0\,km\,s$^{-1}$ or 0.1\,km\,s$^{-1}$ linewidths.
Again, the narrower line width is not consistent with the original
derivation of the temperature profiles \citep{JuvelaYsard2011} but
gives some insight to clouds with higher line optical depths. 

\begin{figure}
\centering
\includegraphics[width=8cm]{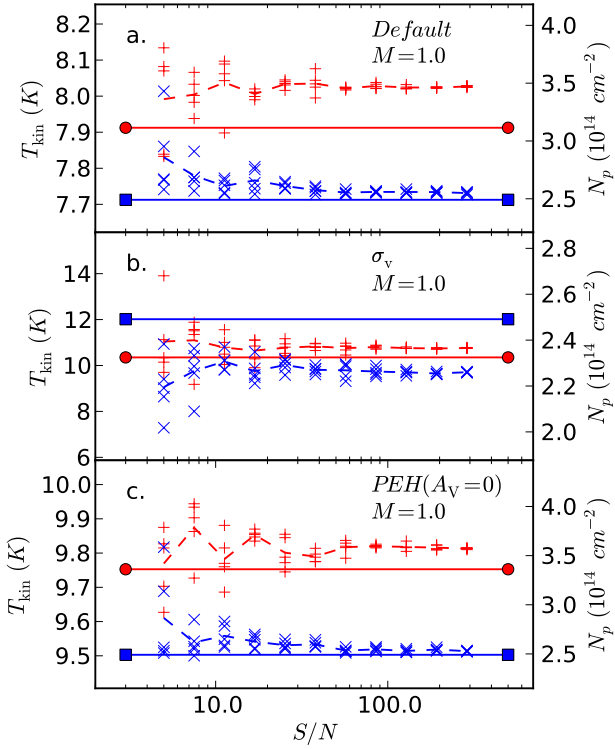}
\caption{
The gas kinetic temperature (plus symbols) and the column density
(crosses) estimated from spectra with different signal-to-noise
ratios. For each S/N ratio, the figure shows results from five
realisations and their average as the dashed lines. The horizontal
lines indicate the true values in the models. The three frames
correspond to the default model ($def$'), the $\sigma_{\rm V}$-model,
and the $PEH$ model, respectively, of the 1.0\,$M_{\rm sun}$ cloud.
}
\label{fig:SNR_test}%
\end{figure}

Figure~\ref{fig:SNR_test} shows the estimated kinetic temperatures and
column densities in relation to their true values in the model
clouds. The spectra were calculated using a pencil beam. The S/N
covers a range from $\sim$3 to 300 in multiplicative steps of 1.5. For
each S/N ratio we show the results from five realizations of spectra,
the dashed lines corresponding to their mean value. 
The results from the $M=0.5\,M_{\sun}$ and the $\sigma_{\rm
V}$=0.1\,km\,s$^{-1}$ were qualitatively similar.

\subsection{Models of high optical depth}

In the default 1\,$M_{\rm sun}$ model the peak optical depths are
$\tau_{\rm (1,1)}$=0.7 and $\tau_{\rm (2,2)}$=0.013. In the
0.5\,$M_{\rm sun}$ cloud these increase to 1.2 and 0.04,
respectively. When the turbulent linewidth was decreased from
1.0\,km\,s$^{-1}$ to 0.1\,km\,s$^{-1}$, the values were naturally
higher, 4.3 and 0.33 for the (1,1) and (2,2) lines, respectively.
However, as the satellite lines continue to be mostly optically
thin, this is not a big problem for the analysis of the spectra. 

\begin{figure}
\centering
\includegraphics[width=8cm]{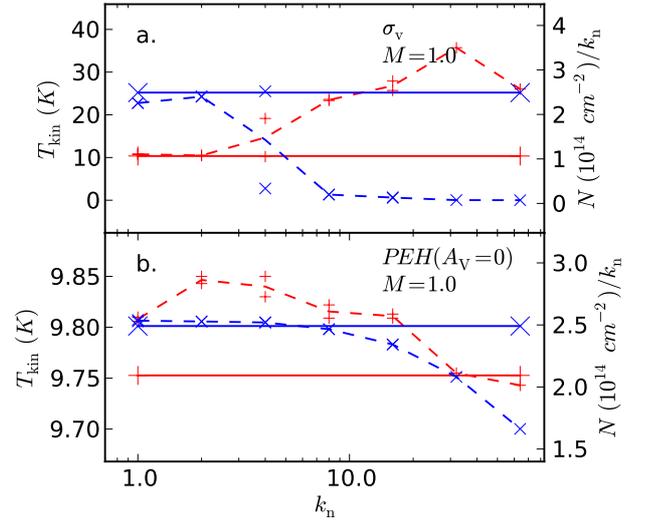}
\caption{
The gas kinetic temperature (plus symbols) and the column density
(crosses) when the densities of the 1.0\,$M_{\rm sun}$ model are
scaled with factors $k_{\rm n}$ in the range of 1-64. The column
densities have been normalized with $k_{\rm n}$. The solid lines
indicate the true values and the dashed lines are the values
deduced from the synthetic observations. The frames correspond to
different temperature distributions as marked in the upper right hand
corner of each frame.
}
\label{fig:n_test}%
\end{figure}

To determine when and how the errors increase to significant levels,
we took the 1.0\,$M_{\rm sun}$ clouds and calculated a series of
models where the densities were multiplied with constant factors
$k_{\rm n}$ between 1 and 64. The temperature profiles are taken from
the original $\sigma_{\rm v}$ and $PEH$ models
(Fig.~\ref{fig:M1.0_sigma1.0}) and are not changed\footnote{We
are investigating the effects of different temperature profiles in
general. However, we note that for high $k_{\rm n}$ the gas kinetic
temperature should approach the dust temperature and $T_{\rm kin}$ is
be expected to decrease towards the core centre \cite[see
e.g.][]{Crapsi2007}. The situation could change only if the core is
heated internally.}
Therefore, the
changes will be only because of the increase of the density and the
column density. Figure~\ref{fig:n_test} shows the resulting $T_{\rm
kin}$ and $N$({\sl para}-NH$_3$) estimates resulting from observations
with a pencil beam towards the cloud centre. As expected, the
estimates show some bias when the column densities are much higher
than for the original 1\,M$_{\sun}$ Bonnor-Ebert spheres. In the $PEH$
model the $T_{\rm kin}$ estimates continue to be very accurate and
even in column density the error rises to $\sim50$\% only when the
density is some 50 times the original values. In this model, the real
temperature is quite constant within the central regions and
apparently the optical depths are still not high enough so that the
spectra would react to the warm but low density surface layers. The
changes were more dramatic in the $\sigma_{\rm v}$-model where the
results were essentially wrong after $k_{\rm n}\sim$4. In this model
the optical depths were higher to start with, because of the narrow
linewidth at the centre.

\begin{figure}
\centering
\includegraphics[width=8cm]{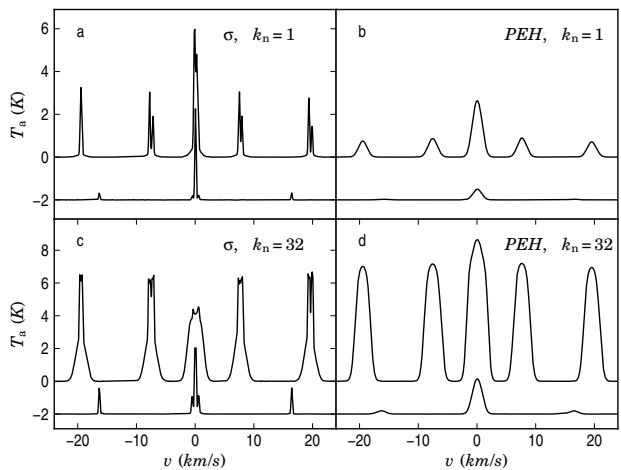}
\caption{
%
%
Comparison of the original spectra from the one solar mass
$\sigma_{\rm v}$ and the $PEH$ models and the spectra obtained after
the densities have been scaled by a factor of $k_{\rm n}$=32. For the
plotting, the (2,2) spectra have been scaled by a factor of 3 (upper
frames) or 0.5 (lower frames) and shifted by -2\,K.
}
\label{fig:spectra_2}%
\end{figure}

Figure~\ref{fig:spectra_2} shows the actual spectra calculated for
$k_{\rm n}$=1 and $k_{\rm n}$=32. In the $\sigma_{\rm v}$-model, the
main group becomes strongly self-absorbed and, because the kinetic
temperature is higher in the centre than at the surface, the intensity
of the satellite lines is higher than for the main group. This is, of
course, a situation that cannot be accounted for in the analysis
assuming a homogeneous medium.

\section{Discussion}  \label{sect:discussion}

\cite{Tafalla2004} examined the validity of the kinetic temperature
estimates from NH$_3$ using isothermal models. In the present study,
we calculate the mass averaged kinetic temperatures in model clouds
with realistic temperature profiles. The values are compared to the
kinetic temperature estimates derived with a standard LTE method,
using the ammonia spectra `observed' from these models.

The estimated kinetic temperatures were generally found to be very
accurate. In the case of model clouds with masses 0.5\,$M_{\sun}$ and
1\,$M_{\sun}$, the difference between the temperature derived
from the NH$_3$ spectra and the real mass-averaged temperature was of
the order of 0.1\,K, only rarely approaching 1\,K. Similarly, the
estimated column densities of para-NH$_3$ remained correct to within
$\sim$10\%.

In one of the models the infall velocity produced two velocity
components whose presence, in the case of large turbulent
linewidths $\sigma_{\rm V}=1.0$\,km\,$^{-1}$, was not immediately
obvious from the line profiles. The analysis completed using a
single gaussian component still led to only $\sim$1\,K error.
A fit with two velocity components resulted in twice as high errors
in both the temperature and the estimated column density
(Fig.~\ref{fig:M1.0_sigma1.0}). When the turbulent line width was
reduced in the cloud core so that the velocity components became
clearly visible in the (1,1) profiles, the error of a single
component fit increased to $\sim$1.5\,K and the column density was
overestimated by 20\% (Fig.~\ref{fig:M1.0_sigma0.1}). However, this
time the two component fit did recover the correct results with
less than 10\% errors. This increases the general confidence in the
results of obtained from ammonia spectra observed with adequate
resolution. It appears unlikely that the effect of hidden features,
e.g. of multiple velocity components, would be amplified in the
$T_{\rm kin}$ and $N$ estimates.

When the signal-to-noise of the (2,2) spectra is decreased below
S/N$\sim$10, the temperature errors increased in our tests to
$\sim$1\,K. The results were not significantly biased so that even
for a set of sources observed at low S/N, the mean properties
should be relatively trustworthy. The sensitivity to noise
naturally depends on the width of the lines. 

When the column density is increased much beyond that found in our
basic models, the (1,1) becomes optically thick and there are no
longer guarantees of the accuracy of the results. In particular, the
lines might no longer probe the conditions at the centre of very dense
cores. We carried out one test with an ad hoc scaling of the
densities. The behaviour was found to depend very much on the line
widths (affecting the initial line opacity) and the temperature
structure of the cloud. In the case with a small turbulent linewidth,
$\sigma_{\rm V}=0.1$\,km$^{-1}$\,s$^{-1}$, completely wrong results
were already encountered with densities a few times higher that those
of an 1\,$M_{\rm sun}$ Bonnor-Ebert sphere. If the fractional
abundance were larger than the 10$^{-8}$ assumed in our models,
observations of normal hydrostatic cores could be affected. However,
in starless cores the values are not likely to be much higher
\citep{Hotzel2001, Tafalla2002, Crapsi2007}.

\begin{figure}
\centering
\includegraphics[width=8cm]{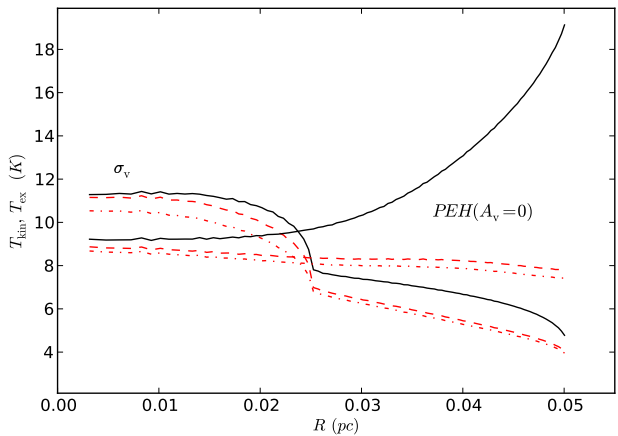}
\caption{
Comparison of temperatures in the $1.0\,M_{\sun}$ cloud models
$\sigma_{\rm v}$ and $peh$. Shown are the radial profiles of the
kinetic temperature (solid lines) and the excitation temperatures of
the (1,1) (dashed line) and (2,2) (dash-dotted line).
}
\label{fig:tex}%
\end{figure}

\cite{Tafalla2004} found that the excitation temperatures of the (1,1)
and (2,2) lines varied by a factor of several. 
This variation depends on several factors, the cloud optical depth,
the volume density, and the variation of the kinetic temperature.
Figure~\ref{fig:tex} shows the situation in those of our
1.0\,$M_{\sun}$ models where the difference in the radial kinetic
temperature profiles is the largest. In the $\sigma_{\rm V}$ model the
kinetic temperature increases inwards and, in combination of higher
volume density and enhanced photon trapping, the excitation
temperatures at the cloud centre are three times as high as at the
cloud boundary. Towards the centre also the difference between the
excitation temperatures of the (1,1) and (2,2) transitions becomes
more noticeable. On the other hand, in the $PEH$ case the outward
increasing kinetic temperature helps to keep the excitation conditions
almost constant. Therefore, it is not surprising that the kinetic
temperature estimated from the observed spectra remained accurate even
when the density was increased (Fig.~\ref{fig:n_test}).

Ammonia was found to be a very reliable tracer of the real mass
averaged gas temperature within the modelled Bonnor-Ebert
spheres. The temperature at the centre of a cloud could still differ
from this number by several degrees. In our models this difference was
at most $\sim$1\,K. However, if the signal from very centre becomes
weaker because of a lower abundance combined with a low kinetic and
excitation temperatures, the difference can be more significant.

In this paper we have assumed the ammonia abundance to be constant
as a function of the cloud radius \citep[see e.g.][]{Tafalla2006}.
If the abundance varies, NH$_3$ is likely to provide accurate
estimates of the average temperature weighted not only by the mass
but also by the fractional abundance. If the ammonia abundance
increases towards the centre \citep[e.g.][]{Tafalla2002} the result
will be a better estimate of the central core temperature. In cores
denser than considered here the situation may be reversed as even
NH$_3$ becomes depleted in the very centre of the core
\citep{Aikawa2005}.
As an illustration of the associated uncertainties, we considered
the $\sigma_{\rm v}$ model with 1.0\,M$_{\sun}$, varying the
abundance between 10$^{-7.5}$ and 10$^{-8.5}$ as a linear function
of the logarithm of the radius. One obtains an estimate of $T_{\rm
kin}$=10.49\,K when the abundance increases inwards and 10.59\,K
when the abundance increases outwards. Both are within 0.25\,K of
the value from the previous constant abundance model.
Thus, in this case the effect of the abundance gradients is not
very significant.
However, we will return to this questions in future papers that
discuss self-consistent models of the chemistry and of the thermal
balance.

\section{Conclusions}  \label{sect:conclusions}

We have calculated ammonia emission spectra using non-LTE radiative
transfer modelling and cloud core temperature profiles taken from the
literature. We have carried out analysis of the simulated spectra and
compared the results with the true parameters of the model clouds.
This has led to the following conclusions:
\begin{itemize}
\item For the $\sim$1\,$M_{\sun}$ hydrostatic clouds, 
the kinetic temperatures and column densities derived from the ammonia
observations are accurate to better than 10\% accuracy.
\item The presence of several velocity components can be identified
from the line profiles before they have a significant effect on the 
accuracy of the analysis.
\item The uncertainty of the kinetic temperature estimates increases to
$\sim$1\,K when the signal-to-noise of the (2,2) spectra is decreased to
S/N$\sim$10.
\item In models of higher density, the behaviour depends the kinetic
temperature profile of the cloud. Quite wrong results can already be
encountered with models a few times more opaque than the 1\,$M_{\rm
sun}$ Bonnor-Ebert spheres.
\item Apart from the most opaque cores, the ammonia was found to be a 
very reliable tracer of the mass averaged kinetic temperature. The
temperature at the centre of dense cores could still differ from this
value by several degrees.
\end{itemize}

\begin{acknowledgements}
The authors acknowledge the support of the Academy of Finland Grants
No. 127015, 250741, and 132291.
The authors thank the referee Dr. Neal Evans for helpful comments,
which improved the paper.

\end{acknowledgements}

\bibliography{biblio_v2.0}

\begin{thebibliography}{22}
\expandafter\ifx\csname natexlab\endcsname\relax\def\natexlab#1{#1}\fi

\bibitem[{{Aikawa} {et~al.}(2005){Aikawa}, {Herbst}, {Roberts}, \&
  {Caselli}}]{Aikawa2005}
{Aikawa}, Y., {Herbst}, E., {Roberts}, H., \& {Caselli}, P. 2005, \apj, 620,
  330

\bibitem[{{Bergin} \& {Tafalla}(2007)}]{BerginTafalla2007}
{Bergin}, E.~A. \& {Tafalla}, M. 2007, \araa, 45, 339

\bibitem[{{Crapsi} {et~al.}(2007){Crapsi}, {Caselli}, {Walmsley}, \&
  {Tafalla}}]{Crapsi2007}
{Crapsi}, A., {Caselli}, P., {Walmsley}, M.~C., \& {Tafalla}, M. 2007, \aap,
  470, 221

\bibitem[{{Danby} {et~al.}(1988){Danby}, {Flower}, {Valiron}, {Schilke}, \&
  {Walmsley}}]{Danby1988}
{Danby}, G., {Flower}, D.~R., {Valiron}, P., {Schilke}, P., \& {Walmsley},
  C.~M. 1988, \mnras, 235, 229

\bibitem[{{Galli} {et~al.}(2002){Galli}, {Walmsley}, \& {Gon{\c
  c}alves}}]{Galli2002}
{Galli}, D., {Walmsley}, M., \& {Gon{\c c}alves}, J. 2002, \aap, 394, 275

\bibitem[{{Goldsmith}(2001)}]{Goldsmith2001}
{Goldsmith}, P.~F. 2001, \apj, 557, 736

\bibitem[{{Harju} {et~al.}(2008){Harju}, {Juvela}, {Schlemmer}, {Haikala},
  {Lehtinen}, \& {Mattila}}]{Harju2008}
{Harju}, J., {Juvela}, M., {Schlemmer}, S., {et~al.} 2008, \aap, 482, 535

\bibitem[{{Harju} {et~al.}(1993){Harju}, {Walmsley}, \&
  {Wouterloot}}]{Harju1993}
{Harju}, J., {Walmsley}, C.~M., \& {Wouterloot}, J.~G.~A. 1993, \aaps, 98, 51

\bibitem[{{Ho} \& {Townes}(1983)}]{Ho1983}
{Ho}, P.~T.~P. \& {Townes}, C.~H. 1983, \araa, 21, 239

\bibitem[{{Hotzel} {et~al.}(2001){Hotzel}, {Harju}, {Lemke}, {Mattila}, \&
  {Walmsley}}]{Hotzel2001}
{Hotzel}, S., {Harju}, J., {Lemke}, D., {Mattila}, K., \& {Walmsley}, C.~M.
  2001, A\&A, 372, 302

\bibitem[{{Juvela}(1997)}]{Juvela1997}
{Juvela}, M. 1997, \aap, 322, 943

\bibitem[{{Juvela} \& {Ysard}(2011)}]{JuvelaYsard2011}
{Juvela}, M. \& {Ysard}, N. 2011, \apj, 739, 63

\bibitem[{{Keto} {et~al.}(2004){Keto}, {Rybicki}, {Bergin}, \&
  {Plume}}]{Keto2004}
{Keto}, E., {Rybicki}, G.~B., {Bergin}, E.~A., \& {Plume}, R. 2004, \apj, 613,
  355

\bibitem[{{Kukolich}(1967)}]{Kukolich1967}
{Kukolich}, S. 1967, Physical Review, 156, 83

\bibitem[{{Mathis} {et~al.}(1983){Mathis}, {Mezger}, \& {Panagia}}]{Mathis1983}
{Mathis}, J.~S., {Mezger}, P.~G., \& {Panagia}, N. 1983, \aap, 128, 212

\bibitem[{{Sch{\"o}ier} {et~al.}(2005){Sch{\"o}ier}, {van der Tak}, {van
  Dishoeck}, \& {Black}}]{Schroier2005}
{Sch{\"o}ier}, F.~L., {van der Tak}, F.~F.~S., {van Dishoeck}, E.~F., \&
  {Black}, J.~H. 2005, \aap, 432, 369

\bibitem[{{Tafalla} {et~al.}(2004){Tafalla}, {Myers}, {Caselli}, \&
  {Walmsley}}]{Tafalla2004}
{Tafalla}, M., {Myers}, P.~C., {Caselli}, P., \& {Walmsley}, C.~M. 2004, \apss,
  292, 347

\bibitem[{{Tafalla} {et~al.}(2002){Tafalla}, {Myers}, {Caselli}, {Walmsley}, \&
  {Comito}}]{Tafalla2002}
{Tafalla}, M., {Myers}, P.~C., {Caselli}, P., {Walmsley}, C.~M., \& {Comito},
  C. 2002, \apj, 569, 815

\bibitem[{{Tafalla} {et~al.}(2006){Tafalla}, {Santiago-Garc{\'{\i}}a}, {Myers},
  {Caselli}, {Walmsley}, \& {Crapsi}}]{Tafalla2006}
{Tafalla}, M., {Santiago-Garc{\'{\i}}a}, J., {Myers}, P.~C., {et~al.} 2006,
  \aap, 455, 577

\bibitem[{{Ungerechts} {et~al.}(1986){Ungerechts}, {Winnewisser}, \&
  {Walmsley}}]{Ungerechts1986}
{Ungerechts}, H., {Winnewisser}, G., \& {Walmsley}, C.~M. 1986, \aap, 157, 207

\bibitem[{{Walmsley} \& {Ungerechts}(1983)}]{WalmsleyUngerechts1983}
{Walmsley}, C.~M. \& {Ungerechts}, H. 1983, \aap, 122, 164

\bibitem[{{Young} {et~al.}(2004){Young}, {Lee}, {Evans}, {Goldsmith}, \&
  {Doty}}]{Young2004}
{Young}, K.~E., {Lee}, J.-E., {Evans}, II, N.~J., {Goldsmith}, P.~F., \&
  {Doty}, S.~D. 2004, \apj, 614, 252

\end{thebibliography}

\end{document}